\documentclass[conference]{IEEEtran}

\usepackage{mathtools}
\usepackage{ifpdf}
\usepackage{lipsum}
\usepackage{cite}
\usepackage{color}
\usepackage{graphicx}
\usepackage{lipsum}
\usepackage{amsmath}
\usepackage{listings}
\usepackage{xcolor}
\usepackage[linesnumbered,ruled]{algorithm2e}
\usepackage{algpseudocode}
\usepackage{float}
\usepackage{textcomp}
\usepackage{array}
\usepackage{placeins}
\usepackage{stfloats}
\usepackage{url}
\usepackage{lipsum}
\usepackage{multirow}
\usepackage{soul}
\usepackage[utf8]{inputenc}
\let\labelindent\relax
\usepackage{enumitem}
\usepackage{algorithmicx}
\graphicspath{{./Figures/}}
\usepackage{glossaries}
\usepackage{tikz}
\usepackage{framed}
\newacronym{IPIC}{IPIC}{Intergovernmental Panel on Climate Change}
\newacronym{ITS}{ITS}{Intelligent Transportation Systems}
\newacronym{GHG}{GHG}{Green House Gases}
\newacronym{ROS}{ROS}{Robot Operating System}
\newacronym{V2V}{V2V}{Vehicle-to-Vehicle}
\newacronym{ADAS}{ADAS}{Advanced Driving Assistance System}
\newacronym{V2I}{V2I}{Vehicle-to-Infrastructure }
\newacronym{V2P}{V2P}{Vehicle-to-Pedestrian}
\newacronym{STARS}{STARS}{Scania Truck And Road Simulation}
\newacronym{TCP}{TCP}{Transmission Control Protocol}
\newacronym{AV}{AV}{Autonomous Vehicle}
\newacronym{plexe}{plexe}{(Platooning Extension for Veins}
\newacronym{TraCI}{TraCI}{Traffic Control Interface}
\newacronym{SUMO}{SUMO}{Simulation of Urban Mobility}
\newacronym{ACC}{ACC}{Adaptive Cruise Control}
\newacronym{CLAC}{CLAC}{Control Look-Ahead Control}
\newacronym{OMNeT++}{OMNeT++}{Objective Modular Network Testbed in C++}

\title{\LARGE \bf Simulation-Based Impact of Connected Vehicles in Platooning Mode on Travel  Time, Emissions and Fuel Consumption}

\author{Aso Validi \emph{Student Member, IEEE}, 
Cristina Olaverri-Monreal \emph{Senior Member, IEEE}
\thanks{Johannes Kepler University Linz, Austria; Chair  Sustainable Transport Logistics 4.0. \texttt{\{aso.validi, cristina.olaverri-monreal\}@jku.at}}}

\begin{document}
	
	\IEEEoverridecommandlockouts
	\IEEEpubid{
			\begin{minipage}{\textwidth}\ \\\\\\\\\\\\\\\\\\[13pt] 
				\copyright 2021 IEEE. Personal use of this material is permitted. Permission from IEEE must be obtained for all other uses, in any current or future media, including reprinting/republishing this material for advertising or promotional purposes, creating new collective works, for resale or redistribution to servers or lists, or reuse of any copyrighted component of this work in other works. DOI: 10.1109/IV48863.2021.9575899
			\end{minipage}}
	\maketitle
	\IEEEpubidadjcol



\begin{abstract}
Several approaches have been presented during the last decades to reduce carbon pollution from transportation. One example is the use of platooning mode. This paper considers data obtained from daily trips to investigate the impact of platooning on travel time, emissions of {$\mathrm{CO_2}$}, {$\mathrm{CO}$}, {$\mathrm{NO_x}$} and {$\mathrm{HC}$} and fuel consumption on a road network in Upper Austria. For this purpose, we studied fuel combustion-based engines relying on the extension of the 3DCoAutoSim simulation platform. The obtained results showed that the platooning mode not only increased driving efficiency but also decreased the total emissions by reducing fuel consumption.
\end{abstract}


\section{{Introduction}}
\label{sec:introduction}

According to a report from the Intergovernmental Panel on Climate Change (IPIC), the transport sector accounts for approximately 23\% of global {$\mathrm{CO_2}$} emissions \cite{Edenhofer2014ClimateBy}. In 2019 a 28\% increase was shown, compared to 1990 levels~\cite{pribyl2020addressing}. This development is mainly due to transportation activities that result from increased customer demand and consequent growth of e-commerce~\cite{liu2020mobile}.
Current transportation systems are highly dependent on burning fossil fuels like gasoline and diesel. Therefore, it is crucial to work towards sustainable transportation through alternatives that reduce fuel consumption and the emissions of greenhouse gases (GHG) that cause the Earth's atmosphere to warm~\cite{pribyl2020addressing}.

Further research towards a sustainable transport strategy includes traveling time and origin-destination matrices to improve travel schedules through the analysis of the most convenient routes for a specific trip \cite{gonccalves2014smartphone}. In this context, a reliable delivery time prediction contributes to eliminating accessory customer trips to a pick up location or avoiding additional travel from the side of the delivery company to return the goods to the sender \cite{9346325}.

There exist other technology-based approaches that pertain to the field of Intelligent Transportation Systems (ITS) that can curtail global temperature increases. For example, autonomous delivery robots can overcome challenges related to navigation in an urban environment, including parking and its effect on fuel consumption, fuel costs and {$\mathrm{CO_2}$} emissions \cite{liu2020mobile}. 

The 3DCoAutoSim simulation platform was originally implemented to mimic Vehicle-to-Vehicle (V2V)~\cite{michaeler20173d} communication from a driver-centric perspective~\cite{biurrun2017microscopic} and has been continuously extended to include automated capabilities and reproduce cooperative Advanced Driving Assistance Systems (ADAS). It uses vehicular data connection among multiple simulators and makes it possible to test applications that are additionally based on Vehicle-to-Infrastructure (V2I) or/and Vehicle-to-Pedestrian (V2P) communication \cite{hussein20183dcoautosim, olaverri2018implementation, olaverri2018connection, artal2019}. The 3DCoAutoSim simulator is also linked to SUMO (Simulation of Urban Mobility)~\cite{behrisch2011sumo} for microscopic road traffic simulation and connected to the Robotic Operation System (ROS)~\cite{quigley2009ros}.

Additionally, other simulators such as the simulator Vehicular Network Interface (Veins)~\cite{sommer2011bidirectionally} uses SUMO and provides an extension for evaluation and analysis of platooning systems from a networking a road traffic perspective (Platooning Extension for Veins, PLEXE)~\cite{segata2014plexe}.
Veins also uses the Objective Modular Network Testbed in C++  (OMNeT++)~\cite{OMNeT++Simulator} network simulator.
This combination enables a detailed simulation of wireless communication among vehicles (in particular IEEE 802.11p-based communication) together with realistic mobility patterns. 

Therefore, we built on PLEXE to extend the vehicular communication capabilities of the 3DCoAutoSim platform and connect several delivery vans traveling on the same route via vehicle-to-vehicle communication. As the Traffic Control Interface (TraCI)~\cite{traciSUMO} provides access to SUMO through a Transmission Control Protocol (TCP)-based client/server architecture, we relied on it to link both simulation platforms. We then used real data collected from delivery trips of bio-products in Upper Austria and analyzed the effects of driving in platooning mode on travel time, GHG emissions and fuel consumption.

The remaining parts of this paper are organized as follows: the next section describes related work; the section~\ref{sec:method} presents the methodology adopted to link the different modular components within the 3DCoAutoSim simulation platform and to perform the comparative analyses; the results from the analyses of the defined scenarios are presented in section~\ref{sec:results}; and the section~\ref{discussion} concludes the work and outlines future research.

\section{{Related Work}}
\label{sec:RelateddWork}
There has been a marked increase in research works that target decreased dependence on fossil fuels in the transportation sector. 

We review and summarize in this section platooning-related published works, highlighting the contribution of our work to the state of the art.

By relying on the Scania Truck And Road Simulation (STARS) \cite{HeavyETDEWEB} simulated environment the authors in~\cite{al2010experimental} investigated the effect of heavy duty vehicle (HDV) platooning on fuel reduction.
Results from the study  showed a maximum fuel reduction of 4.7–7.7\% at a speed of 70 km/h that was dependent on the time gap and the number of vehicles in the platoon. The authors reported vehicle weight to be an additional influential factor on the platooning system, with the fuel reduction of 3.8–7.4\% when the lead vehicle was 10 tonnes lighter in contrast to a 4.3–6.9\% fuel reduction when the lead vehicle was 10 tonnes heavier. 

In further studies relying on the simulation of urban mobility SUMO, the effects of platooning on the generation of emissions were investigated in different networks by applying Cooperative Adaptive Cruise Control (CACC) \cite{erdaugi2019emission, silgu2020network}. The authors relied on the combination of OMNet++ and Veins. The road and network simulation tools assumed the HBEFA3 emission model in all vehicles, according to the classification in the Handbook Emission Factors for Road Transport (HBEFA) \cite{HBEFA}.
The studies focused on exploring different penetration rates of CACC varying from 0-100\% in 25\% increments in two scenarios with intersections regulated by fixed and adaptive traffic light control. The maximum GHG emission reduction reported in the study was 20\% \cite{erdaugi2019emission}.

A two-layer Control Look-Ahead Control (CLAC) architecture for fuel-efficient and safe heavy-duty vehicle platooning was proposed and simulated in~\cite{Turri2017CooperativePlatooning} to obtain road information and real time control of the vehicles. Through Dynamic Programming (DP), the optimal speed of the platoon for maximum fuel efficiency was estimated. For real-time control, a predictive control framework was considered. The approach resulted in a maximum fuel savings of 12\% for following vehicles in the platoon.

This represents a selection of the variety of methodologies developed and adopted to investigate platooning-related metrics that pertain to traffic flow, such as fuel efficiency, travel time, and pollutant emissions.
However, studies that use data generated from real road trips are very scarce. Therefore, we contribute with this work to the gap in the body of knowledge by performing research on platooning using a simulation platform with real data sets collected from real trips in Austria.

\begin{figure}[!t]
	\centering
	\includegraphics[width=0.48\textwidth]{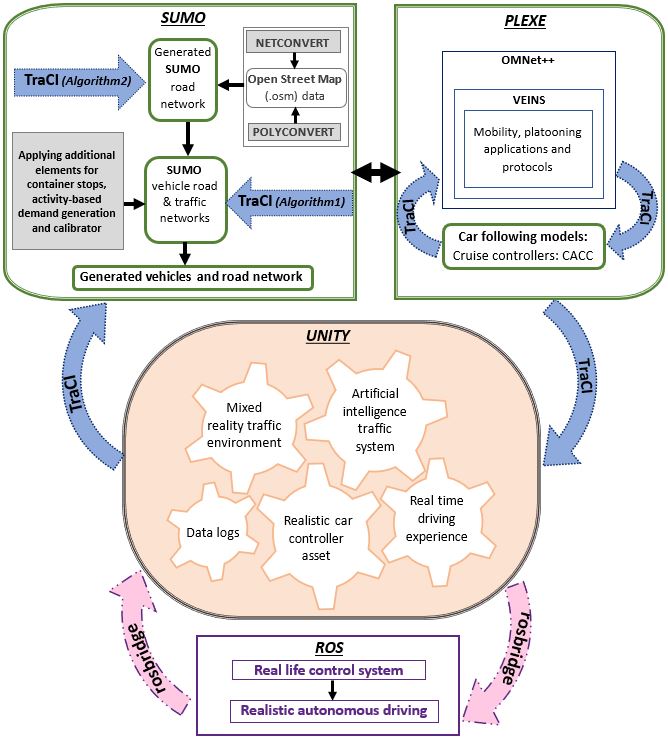}
	\caption{{Overview of the modular components within the ``3DCoAutoSim'' simulation framework}}
	\label{fig:framework}
\end{figure} 


\section{{Methodology}}
\label{sec:method}

As previously mentioned, we extend in this work the ``3DCoAutoSim'' simulation platform with SUMO and the vehicular communication capabilities that are enabled by the Veins network interface and the network simulation OMNet++ to simulate a IEEE 802.11p wireless network-based communication and to design and simulate longitudinal controllers. An overview of the linked components that compose the ``3DCoAutoSim'' simulation framework is presented in  Figure~\ref{fig:framework}. The 3D visualization in the simulator is based on Unity, which makes it possible to simulate a variety of controlled driving environments 
through realistic car controller assets.
UNITY, ROS and SUMO are connected by relying on the rosbridge package \cite{rosbridge} and the TraCI interface.  

To validate the implemented approach, we analyzed the effect of platooning on the dependent variables travel time, total emissions and vehicular fuel consumption. To do this, we compared the data obtained from three delivery vans under the following two scenarios:

\begin{enumerate}
	\item \textbf{Connected} in platooning mode, the vehicles in this scenario being labeled as: Leader.Delivery-Van.1 (LDV1), Follower.Delivery-Van.2 (FDV2) and Follower.Delivery-Van.3 (FDV3). 
\item \textbf{Not connected}, the vehicles in this scenario being labeled as: Delivery-Van.1 (DV1), Delivery-Van.2 (DV2) and Delivery-Van.3 (DV3). 
		
\end{enumerate} 
\begin{table}
	\scriptsize
	\centering
	\caption{Parameters values to setup the simulation scenarios}
	\label{table:simulationparameters}
	\renewcommand{\arraystretch}{1.4}
		\begin{tabular}{|l|l|c|} 
			\hline
			\multicolumn{2}{|c|}{\textbf{Parameters}} & \textbf{Value}~~ \\ 
			\hline
			\multirow{2}{*}{Speed} & min & $5 m/s (18 km/h)$  \\ 
			\cline{2-3}
			& max &  $20 m/s (72 km/h)$  \\ 
			\hline
			\multicolumn{2}{|l|}{Acceleration (max)} & $2.5 m/s^2$  \\ 
			\hline
			\multicolumn{2}{|l|}{Delivery van weight} & $3500 kg$  \\ 
			\hline
			\multicolumn{2}{|l|}{Delivery van length} & $5.94 m$  \\ 
			\hline
			\multicolumn{3}{l}{~}
		\end{tabular}
\end{table}
\begin{table}
	\centering
	\caption{Collected trip data}
	\label{table:gpsparameters}
	\renewcommand{\arraystretch}{1.3}
	\resizebox{\linewidth}{!}{%
		\begin{tabular}{|l|l|} 
			\hline
			\multicolumn{1}{|c|}{ \textbf{Parameters} } & \multicolumn{1}{c|}{\textbf{Explanation} } \\ 
			\hline
			Day & day of tracking (format: T:name of the day) \\ 
			\hline
			Date & date of tracking (format: Y:year, M:month, D:day) \\ 
			\hline
			Time & time of tracking in every second (format: H:hour, M:minute, S:second) \\ 
			\hline
			Latitude n/s & north–south geographic coordinate \\ 
			\hline
			Longitude e/w & east–west geographic coordinate \\ 
			\hline
			Height & elevation above sea level (unit: meter) \\ 
			\hline
			Speed & speed of vehicle at every second (unit: kilometer) \\ 
			\hline
			Heading & compass direction (unit: degree) \\ 
			\hline
			VOX & recorded voice messages \\
			\hline
		\end{tabular}
	}
\end{table}

The relevant parameters to set up the simulation that were common to both scenarios are presented in Table~\ref{table:simulationparameters}. The input that we selected for the gap size parameter in the platooning mode was of $5 m$  \cite{segata2014plexe}. The value corresponded to an average speed of $70.2km/h$ and $71.28 km/h$. 

\begin{figure}[!t]
	\centering
	\includegraphics[width=0.40\textwidth]{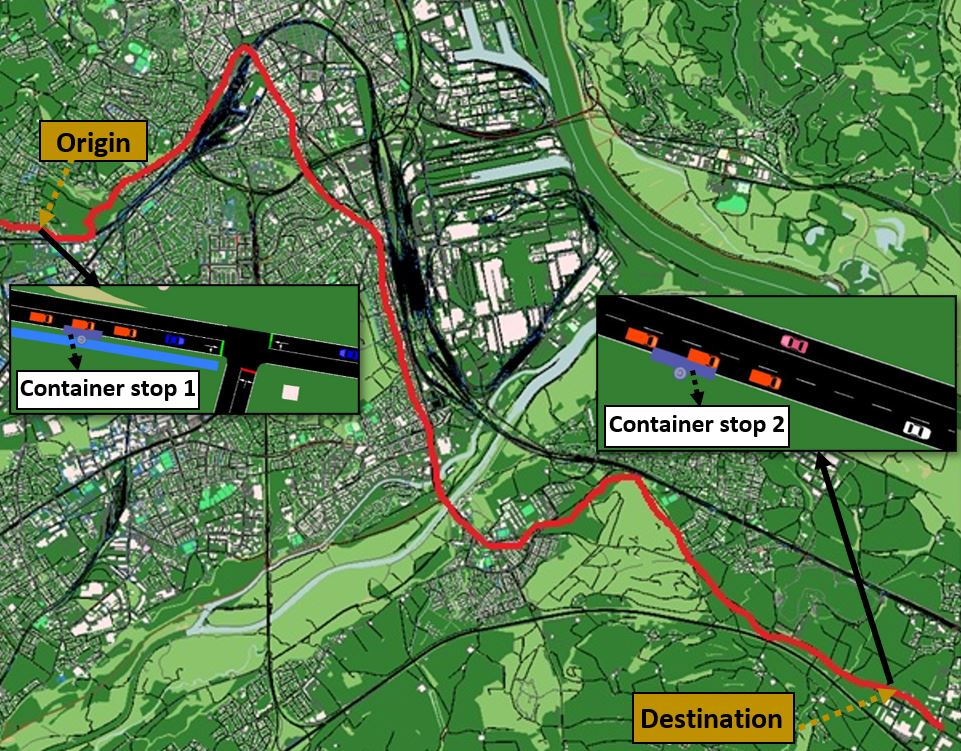}
	\caption{Visualization of the collected travel data showing the driven route ($14 km$) and the pick up and delivery van stops} 
	\label{fig:network}
\end{figure}

The considered route in the scenarios is based on the real GPS data gathered from 12 trips (morning trips at around 11:00 am) with an average travel time and speed of $20min$ and $20m/s$ ($72km/h$) respectively. The trips were performed by the Achleitner Biohof GmbH (Achleitner Organic farm) in the itinerary around Linz in Upper Austria, which are illustrated in the road network in Figure~\ref{fig:network} after the data processing and cleaning procedure. The detailed collected information about the daily trips is presented in Table~\ref{table:gpsparameters}.

\begin{algorithm}[!t]
	\small
	\caption{ \small Applying TraCI for generating the route files and vehicles in the different scenarios} 
	
	\DontPrintSemicolon
	\label{algo:routefiles}
	\SetAlgoLined
	\SetKwData{Left}{left}\SetKwData{This}{this}\SetKwData{Up}{up}
	\SetKwInOut{Input}{input}\SetKwInOut{Output}{output}\SetKwInOut{Define}{define}
	\Input{Road Network, $r$;}
	\Output{Route file, $rou$;}
	\Define{Number of vehicles $i$, Number of time steps $n$, Vehicle types; platooningvType, vanvType}
	
	\SetKwFunction{FMain}{RoutVehGenerate}
	\SetKwProg{Fn}{Function}{:}{End~Function}
	\Fn{\FMain{}}
	{ {random.seed(s)} \\
		{N $\leftarrow$ n} \\
		\Output{.rou.xml (vType, edges, $veh_{id}$)} \; 
		vehNr $\leftarrow$ 0 \;
		\For{i in range(N)}{
			\If{random.uniform(0,1) $<$ each of the three delivery vans:}{
				write output'$<$ generating the vehicle types with different attributes$/>$' format (i,vehNr), file = routes\; 
				vehNr $+= 1$}
		}
	}          	
\end{algorithm} 

\subsection{Traffic simulation implementation}
\label{subsec:sumo}

\begin{figure}[!t]
	\centering
	\includegraphics[width=0.48\textwidth]{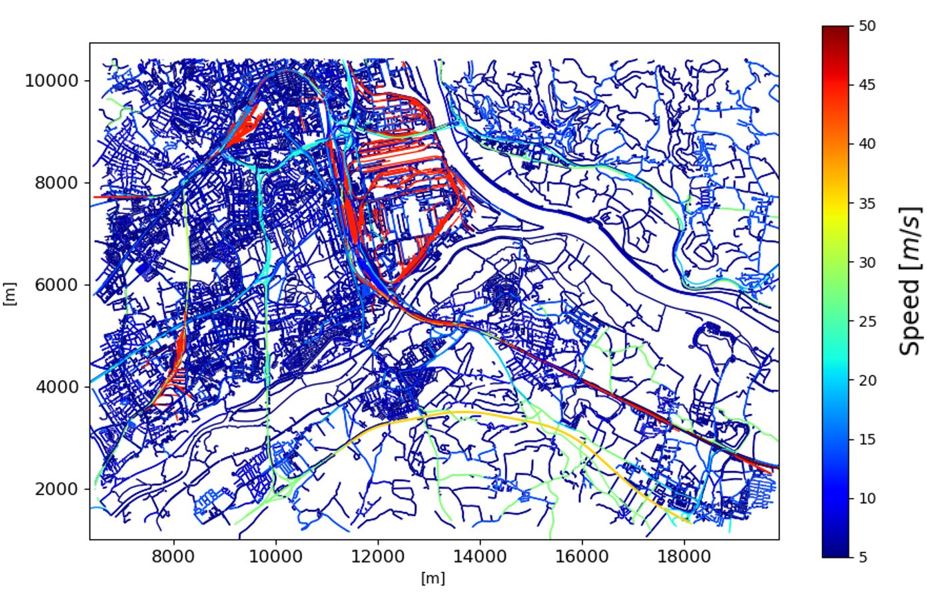}      
	\caption{Visualization of the maximum speed limit in m/s on different routes in the simulated road network} 
	\label{fig:SPEEDMAP}
\end{figure}

To generate the traffic network the map data was imported from Open Street Map (OSM)~\cite{osm} to SUMO. The SUMO application NETCONVERT~\cite{Netconvert} was then executed to import and convert the road network. To interpret the relevant geometrical shapes from the OSM data, an additional typemap-file~\cite{typemap} was utilized, which converted the data to be visualized in the SUMO Graphical User Interface (GUI) through POLYCONVERT \cite{Polyconvert}. 
We also applied the NETEDIT application \cite{NeteditDocumentation} for updating the traffic lights and editing the road network in terms of size and type. In order to have a better overview on the network, we additionally visualized the maximum speed limit on different routes in the road network (Figure~\ref{fig:SPEEDMAP}). 
To replicate reality-close traffic conditions we generated the network demand by relying on statistics from the city of Linz and applying activity-based demand generation using Iterative Assignment (Dynamic User Equilibrium)\cite{DuaiteratSUMO} in 3 iterations.
After these steps, we implemented the vType attribute to characterize the vehicle types from the traffic demand. Each individual vehicle in the simulation was defined by a unique identifier, departure time and planned route.
We additionally adopted TraCI, to be able to retrieve the values of the simulated objects and to manipulate their behavior in real time~\cite{traciSUMO}. Finally, we calibrated the traffic flow in the simulation\cite{CalibratorSUMO}.

Algorithm~\ref{algo:routefiles} describes the process for generating the route files and the traffic demand using TraCI, which used the relevant files in the road network (.osm and .net) as input information. It starts by generating the defined vType and the routes with the related edges on the .rou.xml file. This is then followed by the creation of the vehicles based on their specified probabilities for each of the routes on the .rou file. To generate the different types of vehicles, the algorithm randomly assigns numbers to vehicle IDs. The route file generation follows the defined format by generating all the related attributes, such as type, route ID and departure time. This vehicle creation and definition follows the specified format by generating all the related attributes, such as type, route ID and departure time. To complete all the relevant situations in the simulation, additional elements such as ``container stops''~\cite{containerstop} were defined to mimic loading or delivery locations.

\begin{algorithm}[!t]
	\small
	\caption{\small Applying TraCI for collecting and calculating emissions}
	
	\DontPrintSemicolon
	\label{algo:emission}
	\SetAlgoLined
	\SetKwData{Left}{left}\SetKwData{This}{this}\SetKwData{Up}{up}
	\SetKwInOut{Input}{input}\SetKwInOut{Output}{output}\SetKwInOut{Define}{define}
	\Input{Road Network, $r$;}
	\Output{Emission file; {($GHG$)}  }
	\Define{Emission class; $ec$, Number of vehicles $i$, Number of time steps $n$, Vehicle types}
	
	\SetKwFunction{FMain}{GetVehEmission}
	\SetKwProg{Fn}{Function}{:}{End~Function}
	\Fn{\FMain{$veh_{id}$}}  
	{
		{co2 $\leftarrow$ traci.getCO2Emission $(veh_{id})$} \;
		{co $\leftarrow$ traci.getCOEmission $(veh_{id})$} \;
		{nox $\leftarrow$ traci.getNOxEmission $(veh_{id})$} \;
		{hc $\leftarrow$ traci.getHCEmission $(veh_{id})$} \;
		{\Return EmissionSense} 
		
	}
	\SetKwFunction{FMain}{VehEMISSIONcal}
	\SetKwProg{Pn}{Function}{:}{End~Function}
	\Pn{\FMain{}} 
	{
		{\For{$veh_{id}$ in traci.getIDList():}{
				$veh_{pos}$ $\leftarrow$  traci.getPosition $(veh_{id})$\;
				vehicle $\leftarrow$ Vehicle $(veh_{id}$, $veh_{pos})$ \;
				vehicle.emissions $\leftarrow$ calvehemissions $(veh_{id})$ \;
				vehicles.append(vehicle) \;
				\Return vehiclesemissions}}}
	
\end{algorithm}


\subsection{Vehicle connection definition}
\label{subsec:plexe}

We generated the platooning simulation by modifying the behavior of the vehicles by accessing the related node in OMNeT++, the new behavior being applied to the driven routes. In the OMNeT++ configuring models (.ned and .ini files), we adopted the SimplePlatooningBeaconing class \cite{plexeprotocol}, which extends the base protocol \cite{plexeprotocolbase} for the use of communication protocols. This implemented class is a classic periodic beaconing protocol which sends a beacon every x milliseconds. Furthermore, in order to generate a set of real world simulations with human driven vehicles in combination with platooning, we adopted the  HumanInterferingProtocol \cite{plexeprotocolhuman} application. This class transmits periodic beacons from  human-driven vehicles and makes it possible to generate interfering network traffic. This approach will be part of future studies with real drivers.

The implementation of the simulation for the driven routes from which the data was acquired requires a high  computational capacity that is intensified through the  integration of Veins to link OMNeT++ and SUMO through TraCI. Hence, to run the entire network simulation as smoothly as possible, we generated a few numbers of SUMO vehicles that      were not defined as OMNeT++ nodes and did not have communication capabilities. 
For managing the platoon traffic in the road network, we applied the nCars  and platoonSize configuration parameters from the PlatoonsTrafficManager~\cite{plexetrafficmanager} class. 
The parameters represent the total number of vehicles and the number of vehicles per platoon, respectively. 

To this end, we adopted sinusoidal mode of speed in platoon with oscillation frequency of $0.2Hz$ \cite{segata2014plexe}. We also took advantage of the configuration 
parameters of omnetpp.ini namely ’manager.moduleType’ and ’manager.moduleName’ by setting ’vtypehuman’ to zero (vtypehuman is the parameter definition key for the defined vehicles in PLEXE). 

\begin{figure}[!t]
	\centering
	\includegraphics[width=0.48\textwidth]{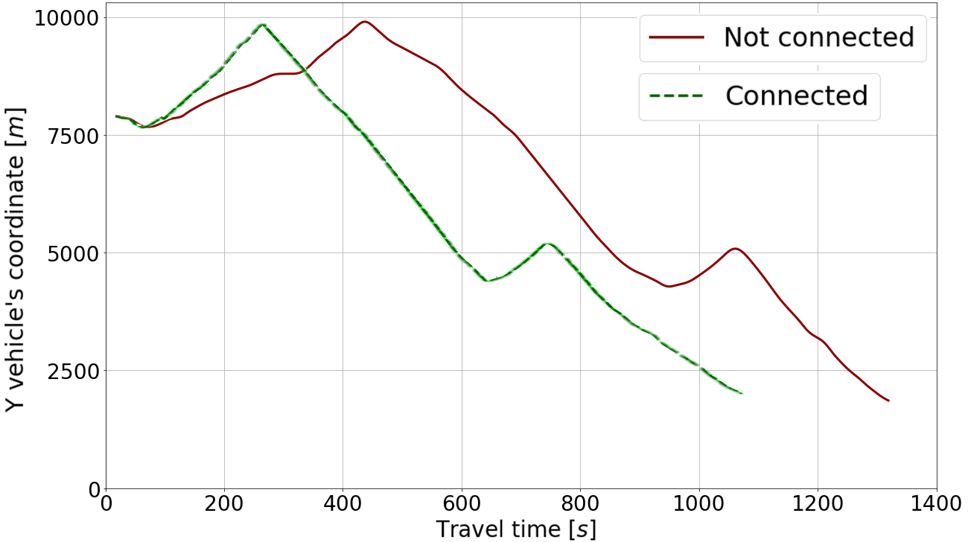}
	\caption{Travel time resulting from comparing the simulated delivery vans in ``connected'' vs ``not connected'' mode}
	\label{fig:traveltime}
\end{figure}

\begin{figure}[!t]
	\centering
	\includegraphics[width=0.48\textwidth]{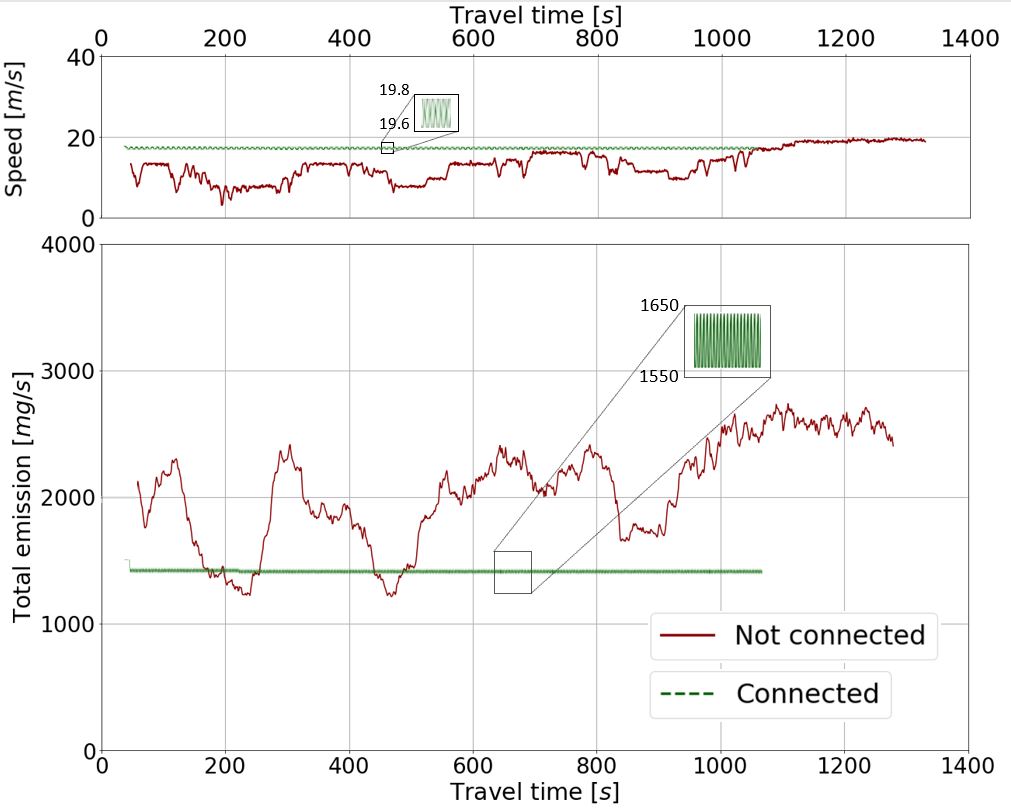}
	\caption{Speed profile (top) and average total emissions (bottom) in ``connected'' vs ``not connected'' mode} 
	\label{fig:emissionboth}
\end{figure}

\subsection{Emissions calculation}
\label{subsec:emission}

To study the total emissions and fuel consumption, we relied on the ``HBEFA3/LDV-D-(EU6)'' emission class that includes the weight and length of the delivery vehicles used~\cite{HBEFA}. 
The calculation of the emissions data is expressed through Algorithm~\ref{algo:emission} using retrievable vehicle variables through TraCI. 
The algorithm uses as input the main road network data files (.rou.xml and .add.xml) with the defined emission class and the vehicle types. It then detects and stores the 
GHG produced by the vehicles ({$\mathrm{CO_2}$}, $CO$, $NOx$, $HC$) in each time step. Based on the vehicle position, the algorithm classifies the gathered data and finally calculates the emissions for each of the vehicles.

\subsection{Comparative analysis description}
To investigate the potential and beneficial effects of driving in platooning mode, we conducted a set of comparative analyses of the use cases defined in section~\ref{sec:method}. To this end, we analyzed the following key metrics: travel time, total emissions and fuel consumption obtained from the simulation.
The total emission analysis was performed by calculating the emission values of the following pollutants {$\mathrm{CO_2}$}, {$\mathrm{CO}$}, {$\mathrm{NOx}$} and {$\mathrm{HC}$} for each use case.

\section{{Results}}
\label{sec:results}

\subsection{Travel time}
The differences in the speed of the ``connected'' compared to the ``not connected'' vehicles (Figure \ref{fig:emissionboth} top) resulted in a reduced travel time of  22.38\% (see Figure \ref{fig:traveltime}). The delivery vans that were not connected needed 1385 seconds each to drive through
the route visualized in Figure~\ref{fig:network}, starting from the ``container stop 1'', defined as origin, to the destination in ``container stop 2''. 

\subsection{Total emissions}

\begin{figure}[!t]
	\centering
	\includegraphics[width=0.44\textwidth]{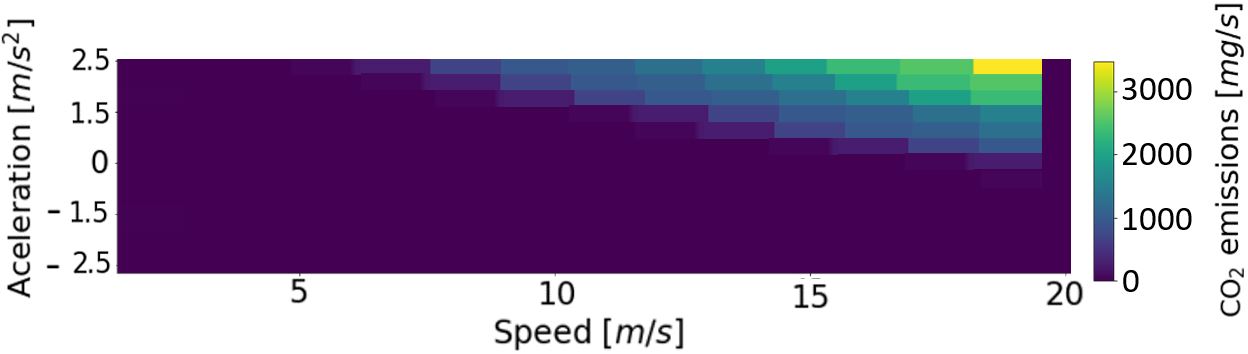}
	\caption{Speed and acceleration effects on {$\mathrm{CO_2}$} emissions in ``not connected'' mode}
	\label{fig:emissionmapnoplatoon}
\end{figure}

\begin{figure}[!t]
	\centering
	\includegraphics[width=0.44\textwidth]{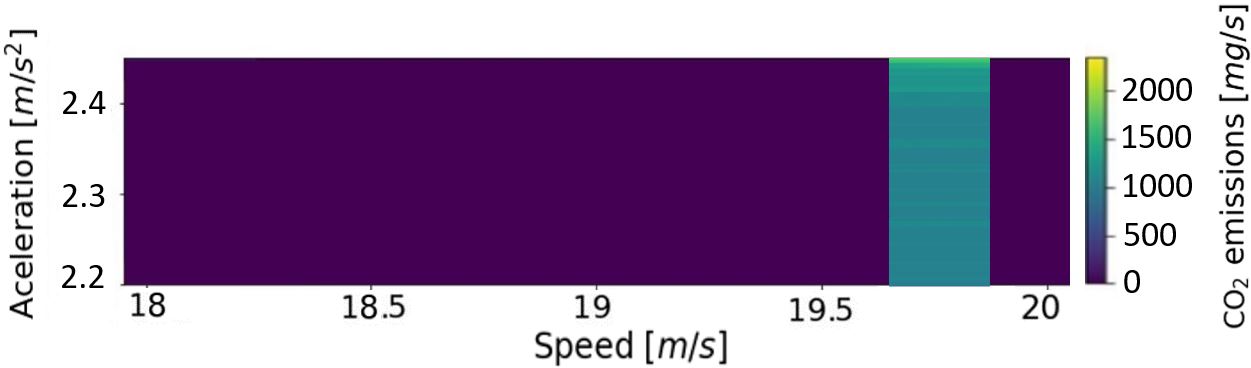}
	\caption{Speed and acceleration effects on {$\mathrm{CO_2}$} emissions  in ``connected'' platooning mode}
	\label{fig:emissionmapplatoon}
\end{figure}

The average emission values obtained from simulating ``connected'' and ``not connected'' scenarios and their related speed profiles are plotted in Figure \ref{fig:emissionboth} bottom and top respectively. In the delivery vans that were not driving in platooning mode, the fluctuations in the total emissions reflect the acceleration and deceleration of the vans throughout the total trip. The driving patterns variation was related to the topology of the road. The higher value of the total emissions in ``not connected'' at the end of the trip (at 1100 s) was due to the higher road speed limit (see speeds in Figure \ref{fig:SPEEDMAP}). 

In the connected platooning mode, all the three vans formed the platoon within a few seconds of the start of the trip, and maintained a steady average speed of $19.8 m/s$ ($71.28 km/h$). The average value of the generated emissions is represented by a flat line with a subtle level of sinusoidal acceleration and deceleration in speed (see section~\ref{subsec:plexe}), which are characteristics of driving in platooning mode (zoomed scale rectangular of Figure \ref{fig:emissionboth}). The values for the total GHG emissions in each of the individual delivery vans are represented in Table \ref{table:summarykeymetrics}.

As carbon dioxide ({$\mathrm{CO_2}$}) is one of the main  GHG that contributes to global warming, we additionally performed a complementary analysis of the generated {$\mathrm{CO_2}$} emissions (Figure~\ref{fig:emissionmapnoplatoon} and Figure~\ref{fig:emissionmapplatoon}). Figure~\ref{fig:emissionmapnoplatoon} shows the relationship between {$\mathrm{CO_2}$} emissions, speed, and acceleration in delivery vans that are not connected. 
As it can be seen, acceleration has a high impact on {$\mathrm{CO_2}$} emissions. The highest value of {$\mathrm{CO_2}$} is emitted at a speed of $19.5 m/s$ ($70.2 km/h$) with an acceleration value of approximately $2.5m/s^2$. Fig~\ref{fig:emissionmapplatoon} shows the relationship between {$\mathrm{CO_2}$} emissions, speed, and acceleration of the delivery vans in  platooning mode. The {$\mathrm{CO_2}$} emissions are, with a value of $1547.34 mg/s$, lower than those in the ``not connected'' mode ($1891.07 mg/s$), due to low variability of the speed  and acceleration metrics (acceleration values between 2.2 and 2.4 $m/s^2$ and speed values between $19.6 m/s$ ($70.56 km/h$) and $19.8 m/s$ ($72 km/h$). The detailed pollutant values of the individual delivery vans for each scenario are presented in Table \ref{table:detailedPollutant}.

\begin{table}
	\centering
	\caption{Emitted pollutant values during the studied trips}
	\label{table:detailedPollutant}
	\renewcommand{\arraystretch}{1.3}
	\resizebox{\linewidth}{!}{%
		\begin{tabular}{|l|l|l|c|l|l|l|l|l|l|} 
			\cline{3-10}
			\multicolumn{1}{l}{} &  & \multicolumn{2}{c|}{\begin{tabular}[c]{@{}c@{}}$Co2$\\\textbf{(cumulated)}\\$mg/s$\end{tabular}} & \multicolumn{2}{c|}{\begin{tabular}[c]{@{}c@{}}\textbf{$NOx$}\\\textbf{(cumulated)}\\$mg/s$\end{tabular}} & \multicolumn{2}{c|}{\begin{tabular}[c]{@{}c@{}}\textbf{$CO$}\\\textbf{(cumulated)}\\$mg/s$\textbf{}\end{tabular}} & \multicolumn{2}{c|}{\begin{tabular}[c]{@{}c@{}}\textbf{$HC$}\\\textbf{(cumulated)}\\$mg/s$\end{tabular}} \\ 
			\hline
			\multirow{3}{*}{\textbf{Connected}} & \textbf{LDV1 } & \multicolumn{1}{c|}{1623.25} & \multirow{3}{*}{\begin{tabular}[c]{@{}c@{}}\textbf{}\\\textbf{SUM=}\\\textbf{4642.04}\end{tabular}} & 1.85 & \multirow{3}{*}{\begin{tabular}[c]{@{}l@{}}\textbf{SUM=}\\\textbf{4.94}\end{tabular}} & 0.196 & \multirow{3}{*}{\begin{tabular}[c]{@{}l@{}}\textbf{SUM=}\\\textbf{0.573}\end{tabular}} & 0.006 & \multirow{3}{*}{\begin{tabular}[c]{@{}l@{}}\textbf{SUM=}\\\textbf{0.014}\end{tabular}} \\ 
			\cline{2-3}\cline{5-5}\cline{7-7}\cline{9-9}
			& \textbf{FDV2 } & 1544.45 &  & 1.67 &  & 0.190 &  & 0.005 &  \\ 
			\cline{2-3}\cline{5-5}\cline{7-7}\cline{9-9}
			& \textbf{FDV3 } & 1474.34 &  & 1.42 &  & 0.187 &  & 0.003 &  \\ 
			\hline
			\multirow{3}{*}{\textbf{Not connected}} & \textbf{DV1 } & \multicolumn{1}{c|}{1991.75} & \multirow{3}{*}{\begin{tabular}[c]{@{}c@{}}\textbf{SUM=}\\\textbf{5673.21}\end{tabular}} & 3.41 & \multirow{3}{*}{\begin{tabular}[c]{@{}l@{}}\textbf{SUM=}\\\textbf{9.99}\end{tabular}} & 0.986 & \multirow{3}{*}{\begin{tabular}[c]{@{}l@{}}\textbf{SUM=}\\\textbf{2.699}\end{tabular}} & 0.037 & \multirow{3}{*}{\begin{tabular}[c]{@{}l@{}}\textbf{SUM=}\\\textbf{0.094}\end{tabular}} \\ 
			\cline{2-3}\cline{5-5}\cline{7-7}\cline{9-9}
			& \textbf{DV2 } & 1883.48 &  & 3.36 &  & 0.871 &  & 0.034 &  \\ 
			\cline{2-3}\cline{5-5}\cline{7-7}\cline{9-9}
			& \textbf{DV3 } & 1797,98 &  & 3.22 &  & 0.842 &  & 0.023 &  \\
			\hline
		\end{tabular}
	}
\end{table}

\subsection{Fuel consumption}

\begin{figure}[!t]
	\centering
	\includegraphics[width=0.48\textwidth]{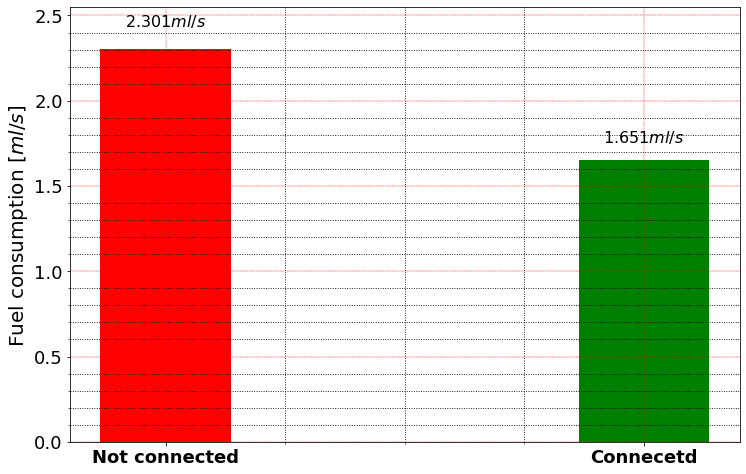}
	\caption{Comparison between cumulative fuel consumption }
	\label{fig:fuel}
\end{figure}

Cumulative fuel consumption results for ``connected'' and ``not connected'' are presented in Figure~\ref{fig:fuel}, the graphic showing a 28.25\% reduction of fuel consumption when the delivery vans are connected and build a platoon. A summary of all the obtained values for ``travel time'', ``total emissions'', and ``fuel consumption'' are presented in Table \ref{table:summarykeymetrics}. 

\begin{table}[!t]
	\centering
	\caption{Key Metrics in the Studied Scenarios}
	\label{table:summarykeymetrics}
	\renewcommand{\arraystretch}{1.2}
	\resizebox{\linewidth}{!}{%
		\begin{tabular}{|l|l|c|l|l|l|l|} 
			\cline{3-7}
			\multicolumn{1}{l}{} &  & \begin{tabular}[c]{@{}c@{}}\textbf{Travel Time}\\$s$\end{tabular} & \multicolumn{2}{c|}{\begin{tabular}[c]{@{}c@{}}\textbf{Total Emissions}\\\textbf{(cumulated)}\\$mg/s$\end{tabular}} & \multicolumn{2}{c|}{\begin{tabular}[c]{@{}c@{}}\textbf{Fuel Consumption}\\\textbf{(cumulated)}\\$ml/s$\end{tabular}} \\ 
			\hline
			\multirow{3}{*}{\textbf{Connected }} & \textbf{LDV1 } & \multirow{3}{*}{\textbf{1075}} & 1625.30 & \multirow{3}{*}{\begin{tabular}[c]{@{}l@{}}\textbf{SUM =}\\\textbf{4647.57}\end{tabular}} & 0.597 & \multirow{3}{*}{\begin{tabular}[c]{@{}l@{}}\textbf{SUM =}\\\textbf{1.651}\end{tabular}} \\ 
			\cline{2-2}\cline{4-4}\cline{6-6}
			& \textbf{FDV2 } &  & 1546.31 &  & 0.557 &  \\ 
			\cline{2-2}\cline{4-4}\cline{6-6}
			& \textbf{FDV3 } &  & 1475.95 &  & 0.497 &  \\ 
			\hline
			\multirow{3}{*}{\textbf{No connected}} & \textbf{DV1 } & \multirow{3}{*}{\textbf{1385}} & 1996.18 & \multirow{3}{*}{\begin{tabular}[c]{@{}l@{}}\textbf{SUM =}\\\textbf{5685.99}\\\end{tabular}} & 0.786 & \multirow{3}{*}{\begin{tabular}[c]{@{}l@{}}\textbf{SUM =}\\\textbf{2.301}\end{tabular}} \\ 
			\cline{2-2}\cline{4-4}\cline{6-6}
			& \textbf{DV2 } &  & 1887.74 &  & 0.772 &  \\ 
			\cline{2-2}\cline{4-4}\cline{6-6}
			& \textbf{DV3 } &  & 1802.06 &  & 0.743 &  \\
			\hline
		\end{tabular}
	}
\end{table}
\section{{Conclusion and Future Work }}
\label{discussion}

To increase the efficiency of transportation, a variety of alternatives has been investigated in the past few decades. Vehicular connectivity and platooning are among the most promising technological ITS solutions, as they reduce fuel consumption and the amount of generated GHG.
In this paper we presented a microscopic simulation-based method to investigate the impact of platooning on travel time, total emissions, and fuel consumption using fuel combustion delivery vans. Based on the results obtained, we conclude that the implemented approach is a feasible solution to investigate the effect of platooning on relevant global warming parameters such as travel time, fuel consumption and the generated emissions from fuel combustion vehicles. Further work in this area will be pursued to collect data involving drivers and electric vehicles in an urban area.

\color{black}
\section*{ACKNOWLEDGMENT}
This work was supported by the FFG project ``Zero Emission Roll Out Cold Chain Distribution 877493'' and the Austrian Ministry for Climate Action, Environment, Energy, Mobility, Innovation and Technology (BMK) Endowed Professorship for Sustainable Transport Logistics 4.0., IAV France S.A.S.U., IAV GmbH, Austrian Post AG and the UAS Technikum Wien. 

\bibliographystyle{IEEEtran}
\bibliography{references}
\end{document}